\documentclass[trackchanges,twocolumn]{aastex701}

\usepackage{booktabs}
\usepackage{placeins}
\usepackage{makecell}

\begin{document}

\title{The \emph{HST}-Hyperion Survey: X-ray AGN Enhancement in the Hyperion Proto-Supercluster }

\author[0009-0008-9352-1572]{Ashlen Coburn}
\affiliation{Institute for Astronomy, University of Hawai'i at M\=anoa, Honolulu, HI 96822, USA}
\affiliation{Department of Astronomy, University of Michigan, Ann Arbor, MI 48109, USA}
\email[show]{acoburn827@gmail.com}

\author[0009-0003-2158-1246]{Finn Giddings}
\affiliation{Institute for Astronomy, University of Hawai'i at M\=anoa, Honolulu, HI 96822, USA}
\email{}

\author[0000-0001-5796-2807]{Derek Sikorski}
\affiliation{Institute for Astronomy, University of Hawai'i at M\=anoa, Honolulu, HI 96822, USA}
\email{}

\author[0000-0001-8255-6560]{Roy Gal}
\affiliation{Institute for Astronomy, University of Hawai'i at M\=anoa, Honolulu, HI 96822, USA}
\email{}

\author[0000-0002-1428-7036]{Brian C. Lemaux}
\affiliation{Gemini Observatory, NSF NOIRLab, 670 N. A'ohoku Place, Hilo, HI 96720, USA}
\affiliation{Department of Physics and Astronomy, University of California, Davis, One Shields Avenue, Davis, CA 95616, USA}
\email{}

\author[0000-0001-6003-0541]{Ben Forrest}
\affiliation{Department of Astrophysical \& Planetary Sciences, University of Colorado, Boulder, 2000 Colorado Ave, Boulder, CO 80309, USA}
\affiliation{Department of Physics and Astronomy, University of California, Davis, One Shields Avenue, Davis, CA 95616, USA}
\email{}

\author[0000-0001-5160-6713]{Emmet Golden-Marx}
\affiliation{INAF - Osservatorio Astronomico di Padova, Vicolo Osservatorio 5, 35122 Padova, Italy}
\email()

\author[0000-0001-9495-7759]{Lu Shen}
\affiliation{Department of Physics and Astronomy, Texas A\&M University, College Station, TX 77843, USA}
\affiliation{Fudan University, Shanghai 200433, People's Republic of China}
\email{}

\author[0000-0002-8209-2783]{Devontae C. Baxter}
\affiliation{Department of Astrophysical Sciences, Princeton University, 4 Ivy Lane, Princeton, NJ 08544, USA}
\email{}

\author[0000-0002-9336-7551]{Olga Cucciati}
\affiliation{INAF - Osservatorio di Astrofisica e Scienza dello Spazio di Bologna, via Gobetti 93/3, I-40129 Bologna, Italy}
\email{}

\author[0009-0003-1792-6539]{Joel Diamond}
\affiliation{Department of Physics and Astronomy, College of Charleston, Charleston, SC 29424, USA}
\email{}

\author[0000-0001-7523-140X]{Denise Hung}
\affiliation{Gemini Observatory, NSF NOIRLab, 670 N. A'ohoku Place, Hilo, HI 96720, USA}
\email{}

\author[0000-0003-2119-8151]{Lori Lubin}
\affiliation{Department of Physics and Astronomy, University of California, Davis, One Shields Avenue, Davis, CA 95616, USA}
\email{}

\author[0000-0001-7811-9042]{Ekta Shah}
\affiliation{Department of Physics and Astronomy, University of California, Davis, One Shields Avenue, Davis, CA 95616, USA}
\email{}

\author[0000-0002-8900-0298]{Sandro Bardelli}
\affiliation{INAF - Osservatorio di Astrofisica e Scienza dello Spazio di Bologna, via Gobetti 93/3, I-40129 Bologna, Italy}
\email{}

\author[0009-0004-4516-0558]{Monica Natalia Isla Llave}
\affiliation{INAF - Osservatorio di Astrofisica e Scienza dello Spazio di Bologna, via Gobetti 93/3, I-40129 Bologna, Italy}
\affiliation{Dipartimento di Fisica e Astronomia ``Augusto Righi'', Universit\`a di Bologna, Via Piero Gobetti 93/2, I-40129 Bologna, Italy}
\email{}

\author[0000-0002-2536-1633]{Hyewon Suh}
\affiliation{Gemini Observatory, NSF NOIRLab, 670 N. A'ohoku Place, Hilo, HI 96720, USA}
\email{}

\author[0000-0001-8776-0958]{Laurence Tresse}
\affiliation{Aix-Marseille Université, CNRS, CNES, LAM, Marseille, France}
\email{}

\author[0000-0003-0680-9305]{Fabio Vito}
\affiliation{INAF - Osservatorio di Astrofisica e Scienza dello Spazio di Bologna, via Gobetti 93/3, I-40129 Bologna, Italy}
\email{}

\author[0000-0002-2318-301X]{Giovanni Zamorani}
\affiliation{INAF - Osservatorio di Astrofisica e Scienza dello Spazio di Bologna, via Gobetti 93/3, I-40129 Bologna, Italy}
\email{}

\author[0000-0002-5845-8132]{Elena Zucca}
\affiliation{INAF - Osservatorio di Astrofisica e Scienza dello Spazio di Bologna, via Gobetti 93/3, I-40129 Bologna, Italy}
\email{}

\begin{abstract}
Dense environments can influence both galaxy evolution and supermassive black hole growth, but whether they enhance or suppress (or both) active galactic nuclei (AGN) activity at high redshift ($z\gtrsim2$) remains uncertain. As one of the most massive structures identified in the early universe, the Hyperion proto-supercluster at $z\approx2.5$ provides a unique opportunity to investigate this connection during an epoch near the peak of both cosmic star formation rate and AGN activity density. To examine the impact of galaxy environment on AGN activity, we measure the fraction of X-ray-selected AGN in Hyperion as a whole, its highly overdense peaks, and its outskirts, comparing these measurements to a coeval field sample. We identify X-ray AGN using the deep \emph{Chandra} COSMOS-Legacy Survey and combine optical/near-infrared (NIR) photometry and photometric redshift information from COSMOS2020 with extensive ground-based spectroscopy and \emph{HST}-grism observations. Redshift information is incorporated statistically using a Monte Carlo framework. We find a higher X-ray AGN fraction in Hyperion ($7.4^{+3.2}_{-1.8}\%$) than in the coeval field ($3.2^{+0.6}_{-0.5}\%$), corresponding to a $\sim2.3\sigma$ difference. Strikingly, this enhancement is concentrated in the overdense peaks, where the X-ray AGN fraction reaches $12.3^{+5.5}_{-3.4}\%$ ($15.0^{+6.9}_{-4.4}\%$ in the most massive peaks versus $6.2^{+11.9}_{-2.1}\%$ in the less massive peaks), while the lower-density Hyperion outskirts ($2.1^{+3.6}_{-1.1}\%$) show no evidence of enhancement relative to the field. These results add to the limited but growing evidence for enhanced AGN activity in high-redshift protostructures, suggesting that the dense environments at these epochs may preferentially trigger AGN during the early stages of cluster assembly.
\end{abstract}


\section{Introduction} \label{sec:introduction}

Galaxy properties vary strongly with environment in the lower-redshift, local universe, from late-types typically found in low-density field regions to the prevalence of early-types in dense galaxy clusters \citep[e.g.,][]{1980ApJ...236..351D}. Gas accretion can regulate the supply of material available to galaxies \citep[e.g.,][]{2017MNRAS.466.3460V}, while environmental processes such as ram-pressure stripping can remove gas from galaxies \citep[e.g.,][]{1972ApJ...176....1G,Boselli_2022, xu2026ram}. At higher redshifts, galaxy interactions increasingly provide a mechanism for redistributing gas \citep[e.g.,][]{2020ApJ...904..107S,  2021A&A...653A.111R, 2022ApJ...940....4S}. Specifically, interactions and mergers can drive gas toward galaxy centers, potentially triggering active galactic nuclei (AGN; e.g., \citealp{2019NatAs...3...48S}). Once triggered, AGN can suppress star formation and contribute to galaxy quenching, working with environmental effects to enhance gas removal \citep[e.g.,][]{fabian1999obscured,2007ARA&A..45..117M}. AGN feedback may also augment environmental processes that remove or limit gas in dense environments \citep[e.g.,][]{2015MNRAS.447..969B}. Enhanced AGN activity in dense environments may therefore influence both host galaxies and the surrounding large-scale structure, motivating measurements of its environmental dependence.

Protostructures at $z\gtrsim2$ allow us to examine this connection during the early stages of cluster assembly \citep[e.g.,][]{2016A&ARv..24...14O, 2017ApJ...844L..23C, 2022Univ....8..554A, 2025A&A...704A.101S}. At this epoch, protostructures are still assembling, with their galaxy populations showing enhanced star formation and stellar mass growth \citep[e.g.,][]{2024MNRAS.528.6934S}, a greater relative abundance of massive galaxies \citep[e.g.,][]{forrest2024environmental,2025ApJ...985...61F,2026A&A...708A.114S}, and elevated rates of galaxy interactions \citep[e.g.,][]{2026A&A...706A..55G} compared to field galaxies at the same epoch. These conditions make protostructures important environments for investigating the connection between supermassive black hole (SMBH) growth and the assembly of large-scale structure.

X-ray observations trace actively accreting SMBHs, including AGN whose emission may be obscured at optical wavelengths \citep[e.g.,][]{hickox2018obscured}. While X-ray surveys can miss highly obscured AGN, sources with X-ray emission below the detection threshold, and some radiatively inefficient AGN that may be identified through radio emission \citep[e.g.,][]{Hickox_2009}, the extensive  \emph{Chandra} coverage of the COSMOS field to moderate depths provides a well-characterized catalog of 4016 X-ray sources, enabling a consistent comparison of X-ray-selected AGN between protoclusters and the coeval field \citep{2016ApJ...817...34M, civano2016chandra}.

This relationship between AGN activity and varying environments has been studied in low-redshift ($z\lesssim 0.30$) galaxy clusters, where luminous AGN were found to be {\it less} common in dense environments compared to the field \citep{2006A&A...460L..23P}. At intermediate redshifts ($z\sim1$), X-ray studies suggest a more complex relationship, with AGN being found outside the densest cluster environments \citep{2009ApJ...700..901K} with evidence for a variety of triggering mechanisms \citep{2017MNRAS.466..496R}. At even higher redshifts ($z\sim1-3$), radio-selected AGN can be reliable signposts of galaxy overdensities, with surveys such as CARLA and COBRA identifying galaxy protoclusters using radio AGN as tracers \citep{Hatch_2013,2013ApJ...769...79W,2017ApJ...844...78P}. At $z\geq2$, the relationship between X-ray AGN activity and environment remains less certain. Studies of individual protostructures have found enhanced X-ray AGN fractions relative to the field, including Spiderweb at
$z=2.16$ \citep{Tozzi_2022}, HS 1700+64 at $z=2.30$
\citep{2010MNRAS.407..846D}, SSA22 at $z=3.09$
\citep{2009MNRAS.400..299L}, and MQN01 at $z\sim3.25$
\citep{2025A&A...694A.165T}, although other studies find no signs of enhancement \citep{2019ApJ...874...54M}. More recently, statistical studies across samples of high-redshift protostructures have also found evidence for enhanced AGN activity in dense environments \citep{2025A&A...704A.101S,2026A&A...714A..12I}.

In this study, we focus on the Hyperion proto-supercluster in the COSMOS field \citep{scoville2007cosmic} at $z\approx2.5$, corresponding to a lookback time of approximately 11.15 billion years. Although smaller components of the structure were identified in earlier studies \citep{diener2015protocluster,chiang2015surveying,casey2015massive,lee2016shadow,wang2016discovery}, \cite{2018A&A...619A..49C} connected these overdensities into a proto-supercluster, estimating a total mass of $\sim4.8 \times 10^{15} M_{\odot}$ and identifying seven overdense peaks. More recent mapping of Hyperion using an expanded spectroscopic and photometric data set has refined the extent and properties of the structure \citep{2026A&A...708A.114S}. As one of the most massive known protostructures, Hyperion provides an opportunity to study AGN activity across a wide range of local overdensities.

We use the multiwavelength photometric, spectroscopic, \emph{HST}-grism, and X-ray observations available in the COSMOS field \citep{Forrest_2024,2025ApJ...985...61F} to compare the X-ray AGN activity in Hyperion to a coeval field sample. We account for host-galaxy stellar mass using spectral energy distribution (SED) fitting and a common stellar mass cut, while propagating redshift uncertainties through Monte Carlo realizations. To determine whether X-ray AGN activity is associated with the local environment during the early stages of cluster assembly, we compare the X-ray AGN fraction between the field, Hyperion as a whole, its highly overdense peaks, and its lower-density outskirts. 

The remainder of this paper is organized as follows. In Section~\ref{sec:data} we describe the photometric, spectroscopic, and X-ray data used in this study. In Section~\ref{sec:methods} we describe the identification of X-ray AGN, environmental classification, MC analysis, SED fitting, and AGN fraction measurements. We present our results in Section~\ref{sec:results} and discuss their details in Section~\ref{sec:discussion}. Throughout this work, we assume a $\Lambda$CDM cosmology with $H_0=70~\mathrm{km\,s^{-1}\,Mpc^{-1}}$, $\Omega_\Lambda=0.73$, and $\Omega_m=0.27$.

\section{Photometric and Spectroscopic Data} \label{sec:data}

To derive the properties of our AGN host galaxies, we perform robust SED fitting with LePHARE using photometry from the COSMOS2020 Classic Catalog v7.0 \citep{2022ApJS..258...11W}, which provides far-ultraviolet to mid-infrared photometry for approximately 1.7 million sources across $\sim2\,\mathrm{deg}^2$ of the COSMOS field. The catalog also provides photometric redshifts with a typical precision of $\sim0.01(1+z)$ for $i<22.5$ and better than $0.025(1+z)$ for $i<25$. To improve redshift accuracy, we include spectroscopic redshifts compiled by the C3VO collaboration \citep{2022A&A...662A..33L, forrest2024environmental}, supplemented by spectroscopy from LATIS \citep{newman2020latis}, CLAMATO \citep{Horowitz_2022}, MOSDEF \citep{2015ApJS..218...15K}, and additional surveys, as well as \emph{HST}-grism redshifts from the \emph{HST}-Hyperion survey \citep{2025ApJ...985...61F}. Redshift quality flags are used to determine whether a spectroscopic, grism, or photometric redshift is assigned to a particular source in our MC analysis (Section~\ref{measurements}). Redshift quality flag and reliability assessment descriptions are available in \citet{2022A&A...662A..33L} and \citet{2025ApJ...985...61F} for the grism sources.

To define the galaxy population we will use to measure the AGN fraction across different environments, we construct a parent galaxy sample from sources with COSMOS2020 photometry following the spatial and IRAC magnitude selection of \citet{2026A&A...708A.114S}. This selection provides a stellar mass-limited sample to $\log(M_\star/\mathrm{M_{\odot}})\ge10$ across the Hyperion region and surrounding field. Across the MC realizations, the resulting parent catalog contains an average of approximately 8,223 sources, including 6,992 (85.0\%) with photometric redshifts, 1,085 (13.2\%) with spectroscopic redshifts, and 146 (1.8\%) with HST grism redshifts. Hyperion and coeval field galaxies are identified from this parent sample as described in Section~\ref{sec:methods}.

X-ray AGN in this study were identified using the \emph{Chandra} COSMOS-Legacy Survey catalog \citep{2016ApJ...817...34M}, which combines new and archival \emph{Chandra} observations covering $2.2\,\mathrm{deg}^2$ of the COSMOS field. Effective exposures are $\sim160$ ks over the central $1.5\,\mathrm{deg}^2$ and $\sim80$ ks over the remaining area, with a limiting full-band ($0.5-10$ keV) flux of $8.9\times10^{-16}\ {\rm erg\ cm^{-2}\ s^{-1}}$ \citep{civano2016chandra}. The X-ray luminosities used throughout this study are not corrected for internal absorption. The COSMOS-Legacy catalog provides absorption corrections based on catalog redshifts, which may differ from those adopted in our analysis to recalculate full-band X-ray luminosities (Section~\ref{measurements}).

\section{Methods} \label{sec:methods}

\subsection{X-ray Matching and AGN Identification} \label{xmatch}
We identified X-ray AGN using the \emph{Chandra} COSMOS-Legacy Survey optical/IR counterparts matched by \citet{2016ApJ...817...34M}. These counterparts were cross-matched to our COSMOS2020-based master spectroscopic and photometric catalog. Of the 4016 X-ray sources in the \emph{Chandra} COSMOS-Legacy catalog, 3582 (89.2\%) had corresponding sources in our master catalog. Unmatched X-ray sources may have ambiguous or undetected optical/IR counterparts, particularly if highly obscured, or may have counterparts identified by \citet{2016ApJ...817...34M} that are absent from our catalog due to its more restrictive source selection criteria. We visually inspected the optical/near-infrared (NIR) counterparts of X-ray sources frequently selected across the MC realizations to assess identification reliability. 

After recalculating the full-band X-ray luminosities using the redshift assigned in each MC realization (Section~\ref{measurements}), we found that the X-ray sources in the Hyperion and coeval field samples spanned $\log(L_{\rm X}/\mathrm{erg\,s^{-1}})=43.17-44.85$ across the realizations. All sources, therefore, remained within the luminosity regime where X-ray emission is expected to be dominated by SMBH accretion rather than stellar processes \citep{hickox2018obscured}, and their classification as X-ray AGN did not change between realizations.

\subsection{Environment Classification} \label{environment_class}
Because high-redshift protostructures have not yet virialized, their density distributions are inhomogeneous and contain highly overdense peaks surrounded by regions of more moderate density. We therefore characterized the galaxy environment using the underlying galaxy overdensity field. We determined the galaxy environments using the Voronoi Monte Carlo (VMC) mapping technique described in detail by \citet{2022A&A...662A..33L} and \citet{2026A&A...708A.114S}. Briefly, the VMC maps used available photometric and spectroscopic redshift information to statistically reconstruct the three-dimensional distribution of galaxies. Within overlapping redshift slices, a Voronoi tessellation divided the two-dimensional projected sky plane into cells surrounding each galaxy, with smaller cells corresponding to regions of higher galaxy density. For each redshift slice, we generated 100 MC realizations incorporating the individual galaxy redshift uncertainties, and the resulting maps were combined across redshift slices to construct a three-dimensional density cube. These realizations constructed the published VMC density maps and are distinct from those used to propagate redshift uncertainties in our analysis (Section~\ref{measurements}). Rather than using galaxy density directly, we characterized environment using the galaxy overdensity relative to the background density at each redshift, expressed as a significance, $n_\sigma$ \citep{2018A&A...619A..49C, 2025ApJ...980..155H, 2026A&A...708A.114S}.

Following \citet{2026A&A...708A.114S}, we identified protostructures as contiguous regions with $n_{\sigma}\geq2.5$, containing at least one overdense peak ($n_{\sigma}\geq4$) and an enclosed mass of $M_{\rm tot}\geq10^{13}\,\mathrm{M_{\odot}}$. Within each protostructure, peaks were defined as contiguous sets of voxels with $n_{\sigma}\geq4$, while the remaining connected voxels ($2.5\leq n_{\sigma}<4$) were classified as outskirts.

\begin{figure}[!t]
\includegraphics[width=\columnwidth]{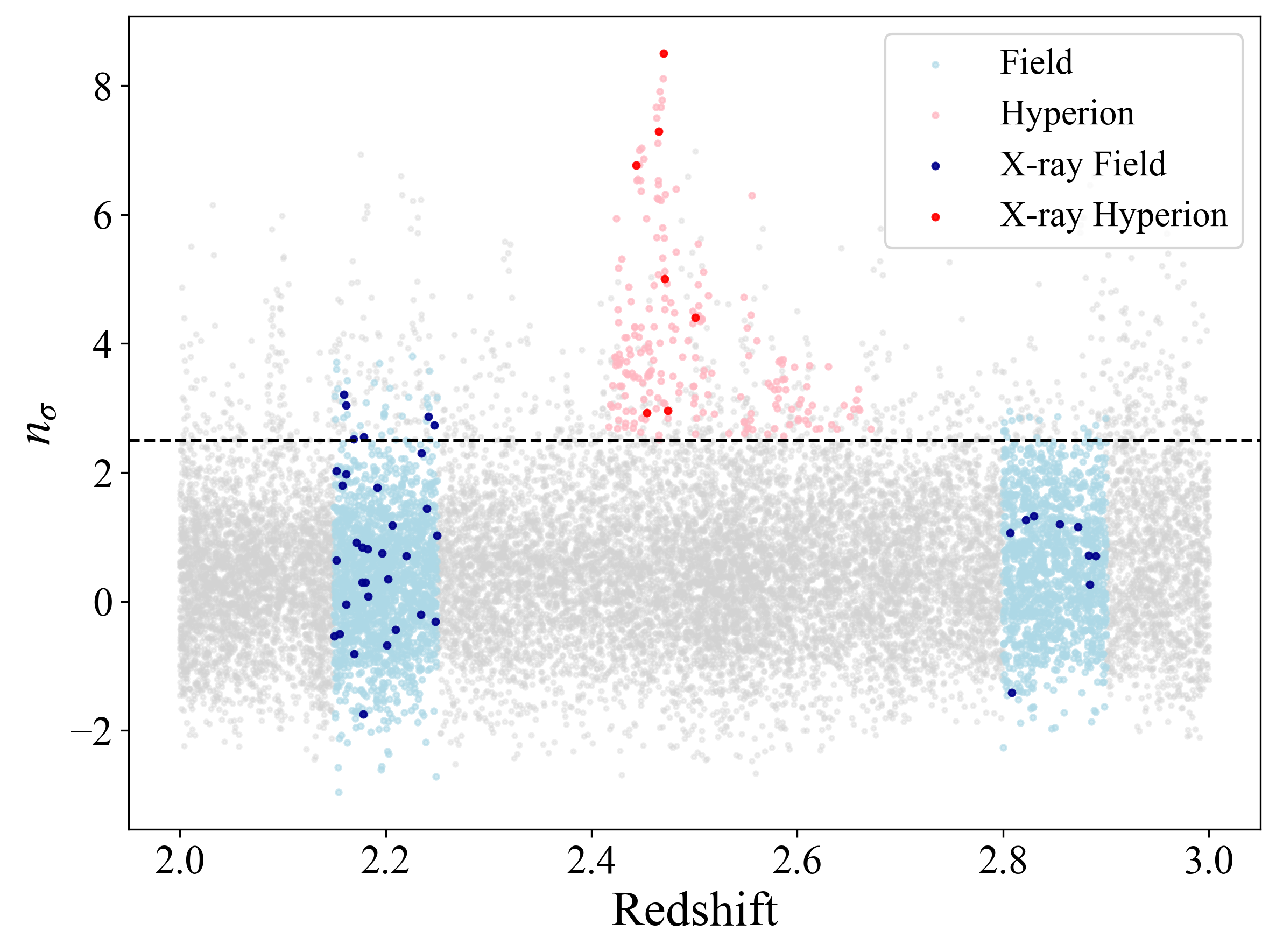}
\caption{Distribution of galaxies in redshift and overdensity ($n_\sigma$) space, using the best available catalog redshift for each galaxy. Galaxies assigned to Hyperion are marked in pink, with corresponding X-ray-identified AGN marked in red. Field galaxies are marked in light blue with corresponding X-ray-identified AGN marked in blue. Unused galaxies are marked in gray (including galaxies outside of the overdensity or redshift range for Hyperion and the field sample). 
\label{fig:n_sigma_vs_z}}
\end{figure}

Galaxies were assigned to Hyperion based on their redshift, sky position, and the corresponding environment classification from the VMC map. The Hyperion sample was further divided into galaxies residing within the overdense peaks and those in the protostructure outskirts. We selected a coeval field sample from galaxies within the same right ascension and declination boundaries as Hyperion but lying in the foreground ($2.15 \leq z \leq 2.25$) and background ($2.80 \leq z \leq 2.90$) redshift intervals. Restricting both samples to the same angular footprint helps ensure similar spectroscopic sampling and spectroscopic redshift fractions, while the field redshift intervals bracket Hyperion to minimize differences in cosmic time and reduce the likelihood that galaxies are scattered between the Hyperion and field samples by redshift uncertainties \citep{2026A&A...708A.114S}. Field membership was determined by whether a galaxy was associated with an identified protostructure, rather than by its local $n_\sigma$ value. Thus, locally overdense galaxies ($n_\sigma\geq2.5$) could remain in the field sample if they were not members of an identified structure. However, only ($1.8^{+0.2}_{-0.2}\%$) of field galaxies were found at $n_\sigma \geq 2.5$ across the MC realizations. The resulting Hyperion and field samples are shown as a function of redshift and local overdensity in Figure~\ref{fig:n_sigma_vs_z}, with X-ray sources indicated by larger and darker points.

\subsection{Monte Carlo Analysis and AGN Fraction Measurements} \label{measurements}

To propagate redshift uncertainties throughout the analysis, the AGN fraction measurements were performed across 100 MC realizations of the galaxy catalog. In each realization, a redshift was assigned to each galaxy using the available spectroscopic, \emph{HST}-grism, and/or COSMOS2020 photometric redshift information. Spectroscopic and grism redshift measurements were assigned according to the reliabilities associated with their quality flags, as described in Section~\ref{sec:data}. When neither measurement was selected, or when no spectroscopic or grism redshift was available, a redshift was randomly sampled from the COSMOS2020 photometric redshift probability distribution \citep{2022ApJS..258...11W}. 

For each MC realization, we fit the 35-band COSMOS2020 photometry at each galaxy's assigned redshift using \textsc{LePhare} \citep{arnouts1999measuring,ilbert2006accurate} and galaxy templates based on the stellar population models of \citet{2003MNRAS.344.1000B}, following \citet{2026A&A...708A.114S}. This generated a unique catalog of derived galaxy properties for each realization.

Because the redshift assigned to an X-ray source could differ from its catalog redshift, we recalculated the full-band X-ray luminosity in each realization using the corresponding luminosity-distance ratio and a $k$-correction assuming $\Gamma=1.4$, consistent with \citet{2016ApJ...817...34M}. Across the MC realizations, an average of 52.6\% of X-ray sources had $|\Delta z|>0.1$ relative to the redshift adopted in the \emph{Chandra} COSMOS-Legacy catalog, with a median $|\Delta z|=0.40$ for these sources. 

Because absorption corrections from \citet{2016ApJ...817...34M} were modeled using the \emph{Chandra} COSMOS-Legacy catalog redshifts, we did not apply them to our recalculated luminosities and instead left luminosities as absorption-uncorrected values. We found that 62.2\% of Hyperion and 54.9\% of field X-ray AGN had catalog-reported absorption correction factors or upper bounds that differed from 1. Among these sources, the mean absorption corrections were comparable between Hyperion and the field, corresponding to luminosity increases of approximately 25\% in both environments. Kolmogorov-Smirnov (KS) and Anderson-Darling (AD) tests returned no statistically significant differences in the absorption-correction distributions between the samples across the MC realizations. We also compared their hardness-ratio distributions as a model-independent test of obscuration, adopting limits for 15.4\% and 28.1\% of Hyperion and field sources, respectively. KS and AD tests again found no statistically significant differences. Although these comparisons cannot rule out differences in highly obscured AGN, the broadly consistent obscuration properties among X-ray-detected sources suggest that our relative comparisons of absorption-uncorrected luminosities would likely remain similar if absorption corrections were applied (i.e., neither sample appears more obscured, at least among the X-ray-detected AGN). 

Because AGN activity depends strongly on host-galaxy stellar mass \citep{2003MNRAS.346.1055K,Aird_2019}, both the Hyperion and coeval field samples were restricted to galaxies with $\log(M_\star/\mathrm{M_{\odot}})\ge10$. We applied this stellar mass cut in each MC realization using the stellar mass derived from that realization's SED fit. Applying the same stellar mass limit to both samples reduced the possibility that differences in AGN fraction were driven by differences in the underlying stellar mass distributions rather than environment. As shown in Section~\ref{sec:results}, we found no statistically significant difference between the stellar mass distributions of the Hyperion and coeval field samples.

To measure a local $n_\sigma$ value, we mapped the assigned MC redshift and sky position of each galaxy into the VMC density cube. Galaxies were assigned to the Hyperion, coeval field, peak, or outskirt samples using the environmental definitions described in Section~\ref{environment_class}. After applying the stellar mass selection, we calculated the X-ray AGN fraction independently in each MC realization as the ratio of X-ray AGN to the total number of galaxies in the sample following \citet{2025A&A...704A.101S}. 

We adopted the median AGN fraction across 100 MC realizations, with MC uncertainties determined from the 16th and 84th percentiles of the AGN fraction distribution. We then combined these uncertainties in quadrature with the uncertainty from the finite number of AGN, calculated assuming binomial statistics following \citet{2011PASA...28..128C}. The AGN enhancement was calculated as the ratio of the median AGN fraction measured in each Hyperion environment to that of the coeval field. We propagated the corresponding upper and lower uncertainties on the Hyperion and field AGN fractions through this ratio to determine the uncertainty on the enhancement.

\section{Results} \label{sec:results}

\begin{figure}
\includegraphics[width=\columnwidth]{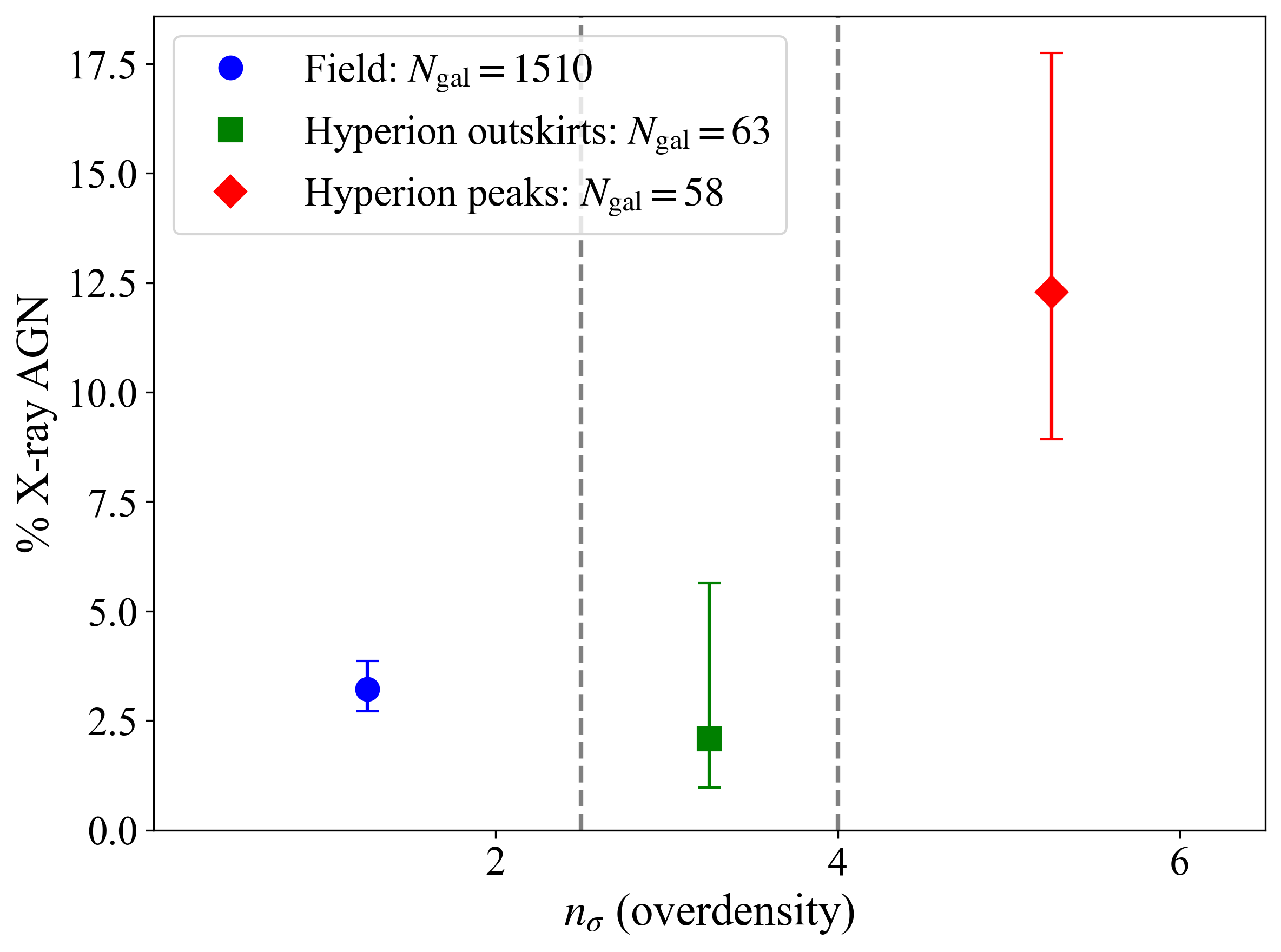}
\caption{X-ray AGN fraction (in \%) as a function of local overdensity in the Hyperion proto-supercluster and coeval field. The galaxy population is divided into three environments: the coeval field (blue), the Hyperion outskirts ($2.5 \leq n_{\sigma} < 4$; green), and the dense Hyperion peaks ($n_{\sigma} \geq 4$; red). Error bars represent the total uncertainties on the AGN fractions, combining the Monte Carlo realizations and Bayesian binomial confidence intervals. The highest X-ray AGN fraction is observed in the dense Hyperion peaks, while the outskirts are consistent with the coeval field within the uncertainties.
\label{fig:xray_agn_fraction_environment}}
\end{figure}

The X-ray AGN fractions measured for each environment are presented in Figure~\ref{fig:xray_agn_fraction_environment} and summarized in Table~\ref{tab:agnfractions}. The coeval field has an X-ray AGN fraction of $3.2^{+0.6}_{-0.5}\%$. In contrast, the full Hyperion sample has an AGN fraction of $7.4^{+3.2}_{-1.8}\%$, a factor of $2.3^{+1.1}_{-0.7}$ higher than the field.

Notably, the enhanced AGN activity is not distributed uniformly throughout Hyperion. Galaxies in the outskirts have an AGN fraction of $2.1^{+3.6}_{-1.1}\%$, consistent with the coeval field. In contrast, galaxies located in the dense Hyperion peaks have an AGN fraction of $12.3^{+5.5}_{-3.4}\%$, a factor of $3.8^{+1.8}_{-1.3}$ higher than the field and a $2.6\sigma$ difference. These results suggest that the enhanced AGN activity is concentrated in the highest-density regions of the protostructure rather than uniformly throughout Hyperion.

\begin{table*}[!t]
\centering
\small
\setlength{\tabcolsep}{8pt}
\caption{X-ray AGN fractions across the Hyperion environments and coeval field. Galaxy and X-ray AGN counts represent the median values across the Monte Carlo realizations, with uncertainties combining the 16th and 84th percentile MC variations and Poisson counting uncertainties in quadrature. AGN fraction uncertainties combine MC variations and binomial uncertainties following \citet{2011PASA...28..128C} in quadrature.}
\label{tab:agnfractions}
\begin{tabular*}{\textwidth}{@{\extracolsep{\fill}}lcccc}
\hline
Environment &
$N_{\mathrm{gal}}$ &
X-ray AGN &
AGN fraction &
Enhancement \\
\hline
Field &
$1510^{+49}_{-51}$ &
$49^{+9}_{-9}$ &
$3.2^{+0.6}_{-0.5}\%$ &
$-$ \\
Hyperion ($n_{\sigma}\geq2.5$) &
$121^{+13}_{-12}$ &
$9^{+4}_{-3}$ &
$7.4^{+3.2}_{-1.8}\%$ &
$2.3^{+1.1}_{-0.7}$ \\
Outskirts ($2.5\leq n_{\sigma}<4.0$) &
$63^{+10}_{-9}$ &
$1^{+3}_{-1}$ &
$2.1^{+3.6}_{-1.1}\%$ &
$0.6^{+1.1}_{-0.4}$ \\
Peaks ($n_{\sigma}\geq4.0$) &
$58^{+9}_{-8}$ &
$7^{+4}_{-3}$ &
$12.3^{+5.5}_{-3.4}\%$ &
$3.8^{+1.8}_{-1.3}$ \\
\hline
\end{tabular*}
\end{table*}

\begin{figure*}[!t]
\centering
\includegraphics[width=0.8\textwidth]{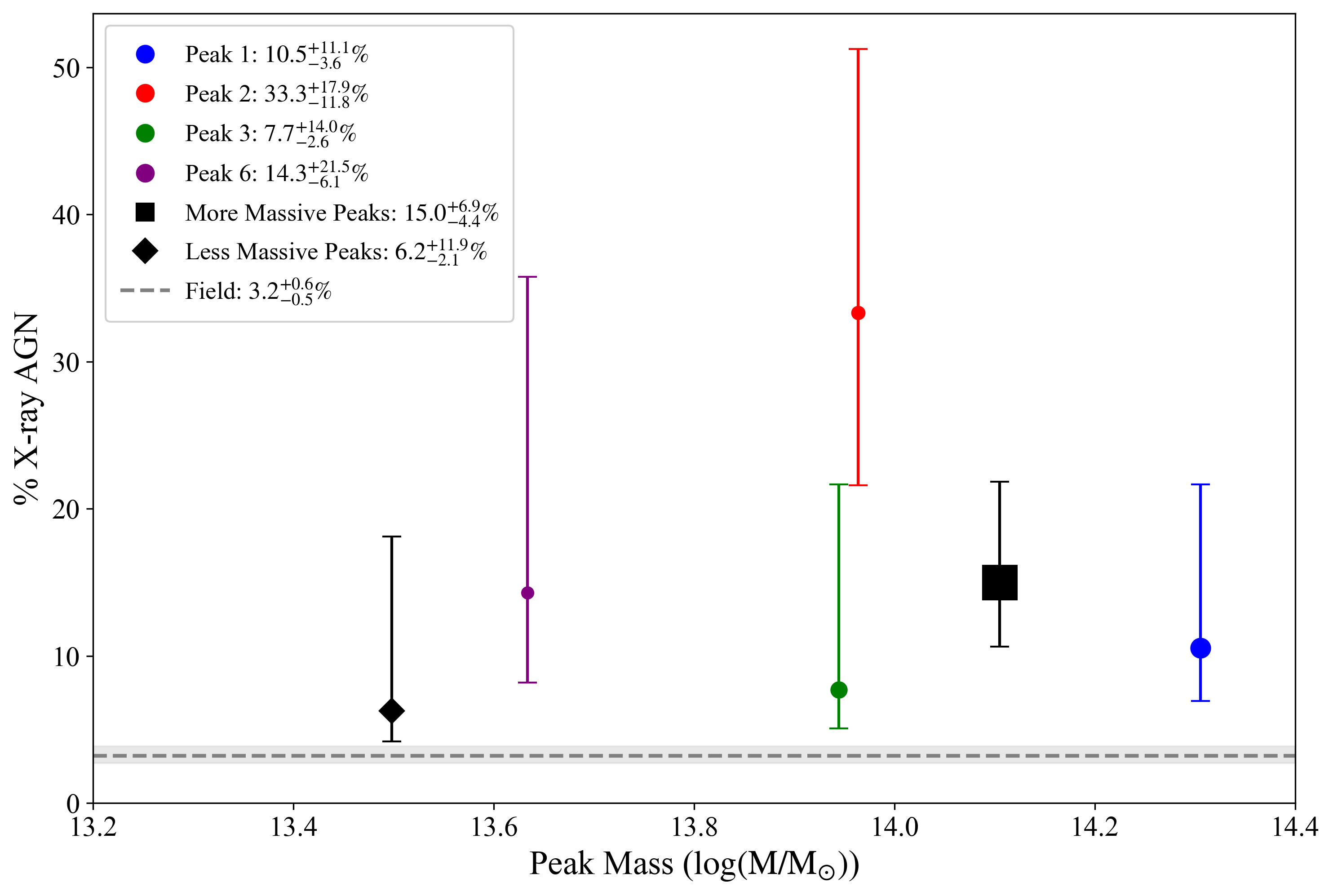}
\caption{X-ray AGN fraction as a function of peak mass for the Hyperion density peaks. Individual points are shown for the four peaks that host X-ray AGN (Peaks 1, 2, 3, and 6). We find X-ray AGN fractions of 
$10.5^{+11.1}_{-3.6}\%$ ($2.0/18.7$) in Peak 1, 
$33.3^{+17.9}_{-11.8}\%$ ($3.0/8.9$) in Peak 2, 
$7.7^{+14.0}_{-2.6}\%$ ($1.1/13.4$) in Peak 3, and 
$14.3^{+21.5}_{-6.1}\%$ ($1.0/7.2$) in Peak 6.  The values in parentheses give the mean $N_{\rm AGN}/N_{\rm gal}$ across the Monte Carlo realizations. Peak masses are taken from \citet{2026A&A...708A.114S}. The size of each point scales with the mean number of total galaxies identified in that peak or peak-mass group across the Monte Carlo realizations. The more massive peaks are Peaks 1--3  with $M_{\rm peak} = 0.88$--$2.02\times10^{14}\,\mathrm{M_{\odot}}$, while the less massive peaks are Peaks 4--7 with $M_{\rm peak} = 0.13$--$0.48\times10^{14}\,\mathrm{M_{\odot}}$. For the two mass groups, the AGN fraction is calculated from the total number of X-ray AGN and galaxies across the included peaks in each Monte Carlo realization, and the point is placed at the mean peak mass of the group. The uncertainty in the mass of each peak is $\sim0.25$ dex \citep{2024MNRAS.528.6934S}. The dashed line and shaded region show the coeval field AGN fraction and its uncertainty.}
\label{fig:agn_fraction_peak_mass}
\end{figure*}

To investigate which Hyperion peaks contribute to the enhanced AGN activity and whether the AGN fraction varies with peak mass, we compare the individual peaks hosting X-ray AGN and the combined more massive and less massive peak populations in Figure~\ref{fig:agn_fraction_peak_mass}, using the peak assignments and masses of \citet{2026A&A...708A.114S}. In the individual peak comparison, we find that Peak~2 has the highest AGN fraction at $33.3^{+17.9}_{-11.8}\%$, while Peaks~1, 3, and 6 have AGN fractions of $10.5^{+11.1}_{-3.6}\%$, $7.7^{+14.0}_{-2.6}\%$, and $14.3^{+21.5}_{-6.1}\%$, respectively. When grouped by peak mass, the three most massive peaks (Peaks~1--3) have a combined AGN fraction of $15.0^{+6.9}_{-4.4}\%$, compared with $6.2^{+11.9}_{-2.1}\%$ for the less massive peaks (Peaks~4--7). Although the measured AGN fraction is higher in the more massive peaks, the large uncertainties, particularly among the less massive peaks, limit our ability to determine whether the X-ray AGN fraction depends on peak mass.

\begin{figure*}[!t]
\centering
\includegraphics[width=0.95\textwidth]
{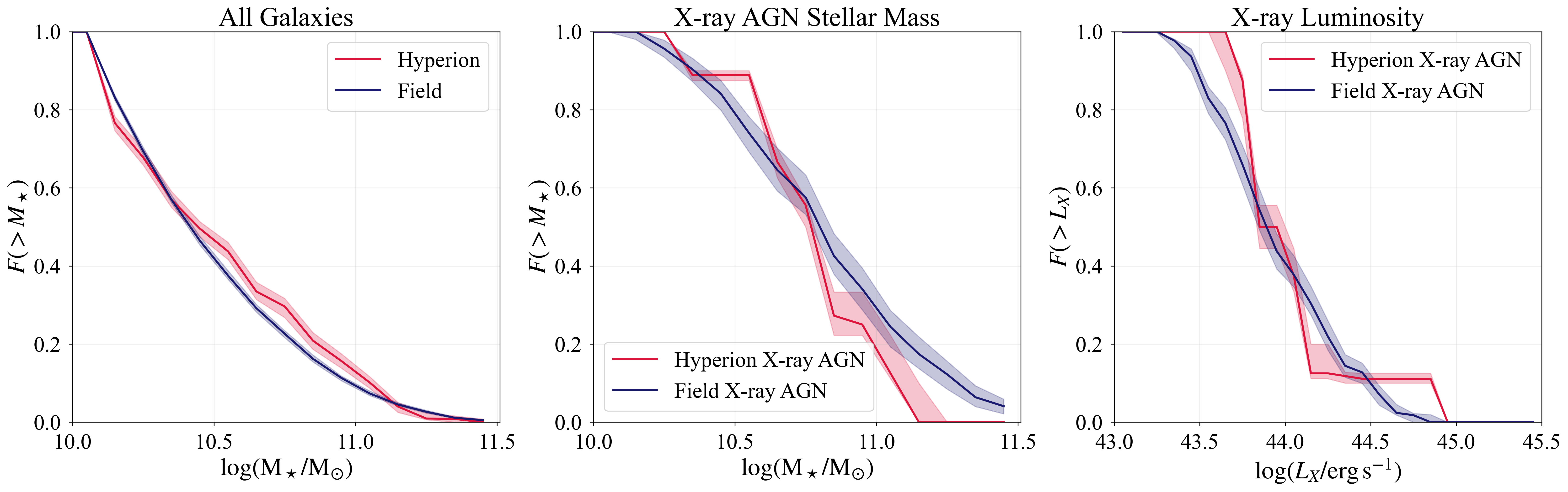}
\caption{Cumulative distributions of stellar mass (left), X-ray AGN host stellar mass (middle), and full-band ($0.5-10$ keV) X-ray luminosity (right) for the Hyperion and coeval field samples. Two-sample KS tests across the 100 Monte Carlo realizations give median $p$-values of $0.236$, $0.784$, and $0.364$, respectively. We find no statistically significant differences between the Hyperion and coeval field distributions in any of the three comparisons.}
\label{fig:cdf_all_xray_mass_and_luminosity}
\end{figure*}

Because AGN incidence depends strongly on host-galaxy stellar mass, we examined whether differences in the stellar mass distributions of Hyperion and the coeval field could contribute to the observed environmental trends, despite the imposition of a stellar mass cut. Figure~\ref{fig:cdf_all_xray_mass_and_luminosity} compares the stellar mass distributions of the mass-selected galaxy populations and X-ray AGN hosts, as well as the full-band absorption-uncorrected X-ray luminosity distributions of the AGN. Two-sample KS tests were performed independently for each of the 100 MC realizations. We find median values of $D=0.094$ and $p=0.236$ for the full galaxy samples, $D=0.222$ and $p=0.784$ for the X-ray AGN host masses, and $D=0.319$ and $p=0.364$ for the full-band X-ray luminosities. AD tests on the stellar-mass distributions show consistent results, with median $A^2=0.907$ and $p=0.136$ for the stellar-mass distribution of the full galaxy samples and $A^2=-0.683$ and $p=0.750$ for the stellar-mass distribution of the X-ray AGN hosts. Neither test provides evidence that the stellar mass distributions differ significantly between Hyperion and the field, although the small number of X-ray AGN limits the statistical power of these comparisons.

\section{Discussion and Conclusions} \label{sec:discussion}
We find an enhanced AGN fraction within Hyperion relative to the coeval field at $z\approx2.5$. This enhancement is concentrated in the densest regions, with the highly overdense peaks exhibiting an enhanced AGN fraction relative to the field while the lower-density outskirts are consistent with the field. This suggests that local environmental conditions within Hyperion may promote SMBH accretion during the early stages of cluster formation.

As discussed in Section~\ref{sec:introduction}, studies of low-redshift galaxy clusters find AGN to be less common in dense environments than in the field \citep[e.g.,][]{2006A&A...460L..23P}, while studies at $z\sim1$ find AGN outside the densest cluster environments \citep[e.g.,][]{2009ApJ...700..901K}. At higher redshifts ($z>2$), studies of AGN activity in protostructures remain limited, but enhanced AGN fractions have been found in individual systems \citep{2009MNRAS.400..299L,2010MNRAS.407..846D,Tozzi_2022,2025A&A...694A.165T} and statistical samples \citep{2025A&A...704A.101S,2026A&A...714A..12I}. Specifically, the $2.3^{+1.1}_{-0.7}$ times higher X-ray AGN fraction in Hyperion is consistent with the $2.4^{+1.9}_{-1.1}$ enhancement found by \citet{2026A&A...714A..12I} for X-ray AGN in Hyperion among submillimeter-selected galaxies. Despite the different galaxy selections, both studies also find no significant differences in the stellar-mass distributions of galaxies or AGN hosts between protostructures and the field, suggesting that stellar mass alone cannot explain the enhancement.

Although the peaks represent the most overdense regions of Hyperion, they are considerably less dense than evolved galaxy clusters at $z\sim1$, and especially those at $z\sim0$. These regions are more comparable to group-like environments, where galaxy mergers and tidal interactions may be more effective at triggering star formation and AGN activity \citep{2009ApJ...700..901K, 2017MNRAS.466..496R}. Studies of Hyperion support this enhanced galaxy growth. Sikorski et al. \emph{in prep.} find higher star formation rates (SFRs) for Hyperion members relative to the coeval field at fixed stellar mass, suggesting that the elevated star formation cannot be explained solely by the presence of more massive galaxies within the proto-supercluster. This star formation is specifically elevated in the Hyperion peaks, which show an order of magnitude increase in the SFR density relative to lower-density environments. Galaxy interactions may contribute to this enhanced activity. In the $z\sim0.9$ Cl1604 supercluster, a majority of X-ray AGN hosts show evidence of recent or ongoing interactions, consistent with observations where interactions contribute to triggering AGN activity at $z\sim1$ \citep{2009ApJ...700..901K, 2013ApJ...768....1M}. Interactions may also be important in Hyperion, where galaxies have close kinematic companions $\gtrsim2.5$ times more frequently than coeval field galaxies \citep{2026A&A...706A..55G}. The elevated interaction rate in Hyperion may contribute to its enhanced AGN activity, potentially with a delay between interactions and SMBH fueling that allows accretion to continue after merger signatures fade \citep{hopkins2008cosmological}.

However, this rapid growth may begin to change in the highest-density regions. \citet{gururajan2025gas} find increasing SFRs but decreasing molecular gas fractions and depletion timescales toward higher densities at the $\sim2\sigma$ level, suggesting rapid galaxy evolution in the densest regions. At the same time, SFRs are highest in the peaks across much of the stellar-mass range but begin to decline above $\log(M_\star/\mathrm{M_{\odot}})\sim10.5$ (Sikorski et al. \emph{in prep.}). The X-ray AGN we identify in Hyperion occupy this higher-mass regime ($10.25 \lesssim \log(M_\star/\mathrm{M_{\odot}}) \lesssim 11.1$) and are preferentially found within the high-density peaks. Enhanced SMBH growth may therefore be occurring alongside declining star formation in the most massive peak galaxies, perhaps suggesting some causal connection. Massive quiescent galaxies have also been identified within dense environments in other high-redshift protostructures \citep{2023ApJ...945L...9I, 2025ApJ...978...17M, 2025arXiv251025024M}.

Several processes could contribute to this transition during the evolution of Hyperion. The shorter gas depletion timescales observed in the denser environments \citep{gururajan2025gas} suggest that rapid gas consumption may limit the fuel available for star formation, and changes in the surrounding environment could decrease the supply of cold gas. Detections of hot intracluster gas in forming clusters at $z>2$ also show that a hot gaseous medium can develop early in cluster assembly through a combination of gravitational heating and AGN feedback \citep{2011A&A...526A.133G,wang2016discovery, Wang_2018,2023Natur.615..809D,2026Natur.649.1130Z}. AGN feedback could also expel gas, although our observations do not establish that AGN activity causes the suppressed star formation observed among the most massive peak galaxies. Hyperion may therefore be observed during a transitional period where rapid star formation and SMBH growth are still occurring while some of the more massive galaxies begin to show quenching and suppressed star formation.

Future work is needed to connect these processes. Analysis of close kinematic companions and the morphologies of the X-ray AGN hosts will help test whether AGN activity is associated with galaxy interactions. Additionally, extending this analysis to more systems, such as those identified by \citet{2025ApJ...980..155H}, and incorporating broader AGN selection methods, will enable more robust measurements of environmental trends and place stronger constraints on AGN activity across different environments at high redshift. 

\begin{acknowledgments}\small

AC acknowledges support from the National Science Foundation Research Experiences for Undergraduates Program at the University of Hawai‘i Institute for Astronomy under NSF grant No.~2548017. AC is grateful to the members of the C3VO collaboration for their guidance, discussions, and support throughout this work.

FV and MIL acknowledge support from 2023 INAF Ricerca Fondamentale Large GO grant.

This work makes use of photometry from the COSMOS2020 catalog, and we thank the COSMOS collaboration for their efforts in obtaining, reducing, and publicly releasing this data. This work also makes use of spectroscopic data compiled by the C3VO collaboration, including observations from the HST-Hyperion survey. We additionally acknowledge the work of the \emph{Chandra} COSMOS-Legacy Survey collaboration, which provides the X-ray source catalog used in this analysis.

This work is based in part on observations made with the NASA/ESA Hubble Space Telescope, obtained from the Space Telescope Science Institute, which is operated by the Association of Universities for Research in Astronomy, Inc., under NASA contract NAS 5-26555. These observations are associated with programs 12328 and 16684. This work is also based in part on observations made with the \emph{Chandra} X-ray Observatory. This work is based in part on observations collected at the European Southern Observatory under ESO programmes 175.A-0839, 179.A-2005, and 185. A0791, as well as work supported by the National Science Foundation under Grant No. 1908422. Supported by the international Gemini Observatory, a program of NSF NOIRLab, which is managed by the Association of Universities for Research in Astronomy (AURA) under a cooperative agreement with the U.S. National Science Foundation, on behalf of the Gemini partnership of Argentina, Brazil, Canada, Chile, the Republic of Korea, and the United States of America. This work is based in part on observations obtained with MegaPrime/MegaCam, a joint project of CFHT and CEA/IRFU, at the Canada-France-Hawaii Telescope (CFHT) which is operated by the National Research Council (NRC) of Canada, the Institut National des Science de l’Univers of the Centre National de la Recherche Scientifique (CNRS) of France, and the University of Hawaii. This research is based in part on data collected at the Subaru Telescope, which is operated by the National Astronomical Observatory of Japan. This work is also based in part on observations made with the Spitzer Space Telescope, which is operated by the Jet Propulsion Laboratory, California Institute of Technology under a contract with NASA.

Some of the spectrographic data presented herein were obtained at the W.M. Keck Observatory, which is operated as a scientific partnership among the California Institute of Technology, the University of California, and the National Aeronautics and Space Administration. The Observatory was made possible by the generous financial support of the W.M. Keck Foundation.

The authors wish to recognize and acknowledge the very significant cultural role and reverence that the summit of Maunakea has always had within the Native Hawaiian community. We are most fortunate to have the opportunity to conduct research that benefits from observations obtained from this mountain.

\end{acknowledgments}

\facilities{CXO, CFHT, GALEX, HST(WFC3), Keck(MOSFIRE), Keck(DEIMOS), Spitzer(IRAC), Subaru(HSC, Suprime-Cam), VISTA(VIRCAM), VLT/ESO}

\software{
Astropy \citep{astropy2022astropy}, Matplotlib \citep{hunter2007matplotlib}, NumPy \citep{harris2020array}, SciPy \citep{virtanen2020scipy}, pandas \citep{reback2020pandas}}

\bibliography{sample701}{}
\bibliographystyle{aasjournalv7}

\end{document}